\documentclass[hyperref]{acmtrans2e}

\usepackage{graphicx}
\usepackage{amsmath}  
\usepackage{psfig}

\newtheorem{definition}{Definition}
\newtheorem{theorem}{Theorem}

\newenvironment{mydef}{\smallskip \begin{definition}}
{\end{definition} \smallskip}

\newtheorem{Ex}{Example}
\newtheorem{Problem}{Problem}

\newtheorem{pro}[theorem]{Proposition}

\firstfoot{}

\runningfoot{}

\title{An Efficient and Flexible Engine for Computing Fixed Points}
\author{Hai-Feng Guo\\University of Nebraska at Omaha
\and
Gopal Gupta\\University of Texas at Dallas}
\begin{abstract}
An efficient and flexible engine for computing fixed-points is  critical 
for many practical applications. In this paper, first, we present 
a goal-directed fixed-point computation strategy
for the logic programming paradigm. Our strategy adopts tabled resolution
(or memoized resolution) to mimic efficient semi-naive bottom-up computation. 
Its main idea is to dynamically  identify and record 
those clauses that will lead to recursive variant calls, and then 
repetitively apply  those alternatives incrementally until the 
fixed-point is reached. Second, we consider those situations in which
a fixed-point contains a large number or even an infinite number of solutions.
In these cases, a fixed-point computation engine may not be efficient enough
or feasible at all. We present a mode-declaration scheme which provides
the capabilities to reduce a fixed-point from a large set of solutions 
to a preferred smaller one, or from an infinite set that is infeasible to
compute to a finite one.  
The mode declaration scheme can be characterized as a meta-level
operation over the original fixed-point. We show the correctness of the 
mode declaration scheme.
Third, the mode-declaration scheme provides a new declarative method 
for dynamic programming, which is typically used for solving 
optimization problems. Using the mode declaration scheme,
there  is  no need  to define the value of  
an  optimal solution recursively, instead, defining  a general solution 
suffices. The optimal value as well as  its corresponding concrete solution 
can be derived implicitly and automatically using a mode-directed 
fixed-point computation engine. Finally, this fixed-point computation 
engine has been successfully implemented in a commercial Prolog system.
Experimental results are shown to indicate that adopting the mode 
declaration scheme improves both time and space performances in solving
dynamic programming problems. 
\end{abstract}








\begin{document}

\begin{bottomstuff}
Author's address: Hai-Feng Guo, Department of Computer Science,
University of Nebraska at Omaha, 6001 Dodge St., Omaha, NE 68182-0500, USA. Email: haifengguo@mail.unomaha.edu\newline
Gopal Gupta, Department of Computer Science,
University of Texas at Dallas, Richardson, TX 75083-0688, USA. Email: gupta@utdallas.edu\newline
\permission
\copyright\ 1999 ACM 0164-0925/99/0100-0111 \$00.75
\end{bottomstuff}
\maketitle

\section{Introduction}

Due    to  their    highly declarative nature and  efficiency, 
Tabled  Logic Programming  (TLP) systems \cite{xsb1,zhou,tals,yap} have
been put to many innovative uses, such as model checking \cite{cav-iv}
and non-monotonic  reasoning \cite{swift},    
A tabled  logic programming system
can be thought of as an engine for efficiently computing fixed-points,
which is  critical for many practical  applications.  A  TLP system is
essential for   extending  traditional LP  system  (e.g.,  Prolog) with
tabled resolution (or memoized resolution). 
The main advantages of tabled resolution are that 
it terminates more  often by computing fixed-points, avoids
redundant computation by memoing  the computed answers, and keeps the
declarative and procedural  semantics  of pure logic programs with 
bounded-size terms consistent.

The main idea  of tabled resolution is never  to compute the same call
twice.  Answers to  certain calls are  recorded in a  global {\em memo
table} (heretofore referred to as a {\em table}), so that whenever the
same call is encountered later, the tabled answers are retrieved and
used instead of being recomputed.  This avoidance of recomputation not
only gains  better efficiency, more importantly, it  also gets  rid of
many   infinite  loops, which often  occur   due to static computation
strategies (e.g.,  SLD resolution \cite{lloyd}) adopted in traditional
logic programming systems.

In this paper, we present a novel tabled resolution strategy for
computing fixed-points in the logic programming paradigm. 
We focus on definite logic programs.
The strategy applies a memoized recursive algorithm 
to mimic efficient semi-naive bottom-up computation. 
Its main idea is to dynamically  identify and record 
those clauses that will lead to recursive variant calls
\footnote{We say two Prolog calls $C_{1}$ and
$C_{2}$ are variant if  there exist substitutions $\phi$ and  $\sigma$
such that  $C_{1} = C_{2}\phi$ and   $C_{2} = C_{1}\sigma$.}, and then 
repetitively apply  these alternatives incrementally until the 
fixed-point is reached. Our tabled resolution scheme allows 
query evaluation to be performed with a single computation 
tree similar to traditional SLD resolution. 
As a result, this novel tabled resolution scheme can be  
easily incorporated into an existing pure Prolog system without
involving any major changes to it. 

\begin{Ex}
Consider the following two programs defining the reachability relation.
The predicate {\tt reach(X,Y)} in the program (a) checks whether 
a node {\tt Y} is reachable from node {\tt X}, while the predicate 
{\tt reach(X,Y,E)} in the program (b) performs the same task but additionally
returns the path from {\tt X} to {\tt Y} as an {\it explanation} (i.e., why
{\tt Y} is reachable from {\tt X}). 
\label{ex:reach}
\end{Ex}
\begin{verbatim}
    :- table reach/2.
    reach(X,Y) :- reach(X,Z), arc(Z,Y).
    reach(X,Y) :- arc(X,Y).
    arc(a,b).    arc(a,c).    arc(b,a).
    :- reach(a,X).

     (a) A Fixed-Point with Finite Number of Solutions



    :- table reach/3.
    reach(X,Y,E) :-
      reach(X,Z,E1), arc(Z,Y,E2), append(E1,E2,E).
    reach(X,Y,E) :- arc(X,Y,E).
    arc(a,b,[(a,b)]).    arc(a,c,[(a,c)]).    arc(b,a,[(b,a)]). 
    :- reach(a,X,P).

     (b) A Fixed-Point with Infinite Number of Solutions
\end{verbatim}

The program in example~\ref{ex:reach}(a), for checking the existence
of reachability, does not work properly in a traditional
Prolog system.  With the declaration of a tabled predicate {\tt reach/2} 
in a tabled Prolog system, it can successfully find the complete set of
solutions due to the fixed-point computation strategy. However, 
there are many situations in which a fixed-point contains 
a large number or even infinite number of solutions, 
which in turn affects the efficiency or completion of 
the computation. 

Consider the reachability program 
shown in example~\ref{ex:reach}(b), where {\tt append/3} is a standard
predicate used to append one list to another. An extra argument is added 
to the predicate {\tt reach/3} to collect the
path that connects its first argument to its second argument. 
However, this extra argument results in 
the fixed-point of the computation becoming infinite. Tabled resolution
now becomes nonterminating, 
since now there are infinite number of paths from {\tt a}
to any node due to the cycle between {\tt a} and {\tt b}.
Similar  problems on evidence construction have been  studied on  
justification in \cite{abhik,just}.  One  reasonable   solution is  
presented in \cite{just}   by asserting the  first evidence into a  
dynamic database for each tabled answer. However, the evidence has 
to  be organized as segments indexed by each tabled  answer.  
That is, an   extra procedure is  required to
construct the full evidence.

To avoid problems of inefficiency or nontermination, due to 
the fixed-point containing a large number or an infinite number 
of solutions respectively, it is often necessary to change 
the original problem so that there is only a small finite-sized
solution set. For the reachability example, it is actually enough 
to find a single simple path to show the evidence of reachability. 
However, generally, it is not only difficult to alter the predicate
definition of {\tt reach/3} to avoid nontermination in order to
have a single simple path for each pair of reachable nodes, 
it also sacrifices the clarity of the original relation. 
In these cases, a fixed-point computation engine may not be efficient enough
or feasible at all. We present a declarative method to get around this problem
by introducing a new mode declaration scheme \cite{mode}. 

In this paper, we present a mode-declaration scheme \cite{mode} in a tabled 
Prolog paradigm which provides the user the capability 
to reduce the fixed-point of a definite logic
program from a large- or infinite-sized solution set to a preferred 
finite-sized one. 
The  method introduces a new  mode declaration for tabled predicates. 
The mode declaration  classifies arguments of a tabled  predicate 
as  indexed or   non-indexed.  Each non-indexed argument can be 
thought of as a function value uniquely determined by the indexed arguments. 
The tabled Prolog system is optimized to perform variant checking 
based only on the indexed arguments during the computation of fixed-points. 
The mode declaration can further extend one of the non-indexed arguments to be
an aggregated value, e.g., the minimum function,  
so that the global  {\em table} will record answers
with the value of that argument appropriately aggregated. 
Thus, in the case of the minimum function, 
a tabled answer can be dynamically replaced by a
new  one with  a  smaller   value during the   computation. 
This new declaration for tabled predicates and  modified procedure for  variant
checking  make it easier for the meta-level manipulation during computation of
fixed-points.

Semantically, the mode declaration scheme can be characterized 
as a meta-level operation over the fixed-point of
the original program. The semantics of a tabled Prolog program
is formalized based on the Herbrand model \cite{van,lloyd} and fixed-point theory, 
whereas the semantics of declared modes is defined as a strict partial
order relation~\footnote{A strict partial order relation
is irreflexive and transitive.} among the solutions.
The mode declarations essentially provide a selection mechanism among
the alternative solutions, thus making fixed-point computation more
flexible. We formally present the semantics of mode declaration 
in a tabled Prolog program, and further show the 
correctness of its operational semantics in tabled resolution. 

The new mode-declaration scheme,  coupled with recursion, provides an
attractive  platform for making dynamic  programming simpler: there is
no  need  to  define  the value  of  an  optimal solution recursively,
instead,  defining the value  of a general  solution suffices. The
optimal value,  as  well as its associated  solution,   will be computed
implicitly  and automatically in a tabled Prolog system that
uses our new mode declaration scheme and modified variant checking. Thus, dynamic
programming problems are solved more elegantly and more declaratively.

We have successfully implemented our tabled resolution as well as the 
mode declaration scheme in the ALS Prolog, a commercial Prolog system.
To implement the mode declaration scheme, no change is required
to the tabled resolution mechanism; therefore, the same idea can also be 
applied to other tabled Prolog systems. Our experimental results show 
that the mode declaration scheme improves both time and space performance 
while solving dynamic programming problems.

The rest of  the paper is  organized as follows: Section~\ref{sec:tlp}
introduces tabled logic programming and our new tabled resolution scheme,
{\em dynamic reordering   of alternatives} (DRA) \cite{tals}, for efficient
fixed-point computation.  Section~\ref{sec:mode} presents a mode declaration 
scheme for tabled predicates, shows how it affects fixed-point
computation, and describes its implementation.
Section~\ref{sec:dynamic} gives a detailed
demonstration of how  dynamic  programming can  benefit from  this new
scheme.  Section~\ref{sec:experiment}
presents the performance results w.r.t. some dynamic programming benchmarks.
Finally,    section~\ref{sec:conclusion}  gives  our conclusions.

\section{Computing Fixed Points via Tabled Resolution}
\label{sec:tlp}

\subsection{Tabled Logic Programming (TLP)}

Traditional logic    programming  systems  (e.g.,   Prolog)  use   SLD
resolution \cite{lloyd} with the following {\it computation strategy}:
subgoals of a resolvent are solved from left to right and clauses that
match a subgoal  are applied in the  textual order they  appear in the
program.   It is   well   known that   SLD  resolution   may  lead  to
non-termination for certain programs, even though  an answer may exist
via  the declarative semantics. That is,  given any static computation
strategy, one can always produce a program  in which no answers can be
found due  to  non-termination   even  though some  answers   may
logically  follow  from the   program.   In case of Prolog,   programs
containing certain  types of  left-recursive  clauses are  examples of
such programs.

Tabled logic  programming  \cite{oldt,xsb1} eliminates such 
infinite loops by extending 
logic programming with tabled resolution. The main idea of tabled 
resolution is to memoize  the answers  to some  calls and
use the memoized  answers   to  resolve  subsequent   variant calls.   
Tabled resolution   adopts  a dynamic  computation   strategy while resolving
subgoals in  the current resolvent against  matched program clauses or
tabled answers. It keeps track of the nature and type of the subgoals;
if  the subgoal in  the  current resolvent  is  a variant  of a former
tabled  call,    tabled answers  are  used to    resolve  the subgoal;
otherwise, program clauses are used following SLD resolution.

The main advantages of tabled resolution are that
a TLP system terminates more  often by computing fixed-points,  avoids
redundant computation by memoing  the computed answers, and  keeps the
declarative and procedural  semantics  consistent for  pure  logic
programs with bounded-size terms. A tabled  logic programming system
can be thought of as an engine for efficiently computing fixed-points,
which is  critical for many practical  applications. 
Tabled  logic programming  (TLP) systems have
been put to many innovative uses, such as model checking \cite{cav-iv}
and non-monotonic  reasoning \cite{swift},    due    to  their    highly
declarative nature and  efficiency.  

In a  tabled  logic programming system,  only  tabled  predicates are
resolved  using  tabled resolution.  Tabled  predicates are explicitly
declared  as follows: \\ 
\centerline{\tt :- table p/n.}\\  
where {\tt p} is a predicate name and {\tt n} is its  arity. 
A global data structure {\em table} is introduced to memoize the 
answers of any subgoals to tabled predicates, and to avoid 
any recomputation.

\subsection{Related Works}

The first tabled resolution scheme, called OLDT \cite{oldt}, was proposed in 1986 
by Tamaki and Sato for avoiding some of the non-termination problems 
during evaluation of definite logic programs. The basic idea of OLDT is to maintain
a global {\em table} data structure to remember the queries seen
so far and their corresponding answers produced so far. 
Such a predicate call that has been  recorded in the  table is referred to as a 
{\it tabled call}.  Any answers to a tabled call will be recorded   in  the  table,  
and referred  to  as  {\it  tabled answers}. OLDT resolution usually maintains
multiple computation trees, in parallel, for computing the fixed-point,
where each computation tree uniquely corresponding to one tabled call.
Subsequently, this work has been extended to SLG resolution \cite{SW94,wff,xsb1} for 
general logic programs with negation. The XSB system \cite{xsb1} is an implementation of 
SLG resolution which supports well-founded semantics \cite{chen,wff}. 
The XSB system has bee implemented atop the Warren Abstract Machine (WAM)
\cite{aitkaci}.  As is well known, compiling to the WAM instruction
set is a standard way of implementing
high performance logic programming systems. The implementation 
of the XSB system involved many non-trivial changes to the WAM, and was
a very substantial engineering effort \cite{xwam}. 

The huge implementation effort needed for implementing OLDT and
SLG can be avoided by choosing
alternative methods for tabled resolutions  that
maintain a single computation tree similar to traditional SLD resolution, rather
than maintaining a {\it forest of SLD trees}.
SLDT resolution \cite{shen,zhou} was the first attempt 
in this direction. The main idea behind SLDT is to 
{\em steal} the backtracking point---using the 
terminology in \cite{shen,zhou}---of the previous tabled call 
when a variant call is found, to avoid exploring the current recursive 
clause which may lead to non-termination.
However, because the variant call avoids applying
the same recursive clause as the previous call, the computation may be
incomplete. Thus, repeated computation of
{\it tabled calls} is required to make up for the lost answers and
to make sure that the fixed-point is complete. SLDT does not propose a
complete theory regarding when a tabled call is completely
evaluated, rather it relies on blindly recomputing the tabled
calls to ensure completeness. SLDT resolution was implemented in early versions
of B-Prolog system. However, recently this resolution strategy has been discarded, 
instead, a variant of DRA resolution \cite{tals} 
has been adopted in the latest version of B-Prolog system \cite{zhou03}.

\subsection{Dynamic Reordering of Alternatives (DRA)}
\label{sec:dra}

The DRA resolution \cite{tals} computes a fixed-point in a  very similar way  as
a goal-directed bottom-up execution of logic  programs \cite{lloyd}. Its main idea  is
to dynamically  identify  {\it looping alternatives} from  the program
clauses, and then repetitively apply  those alternatives until no more
answers can be  found. A looping  alternative refers to a  clause that
matches a  tabled  call and  will   lead to  a resolvent containing  a
recursive variant call. The DRA resolution requires that not only the 
answers to variant calls are tabled, the alternatives leading to variant 
calls are also memorized in the table, which is essentially different from 
all the previous tabled resolution strategies.

\begin{figure}[htb]
\centerline{
\psfig{figure=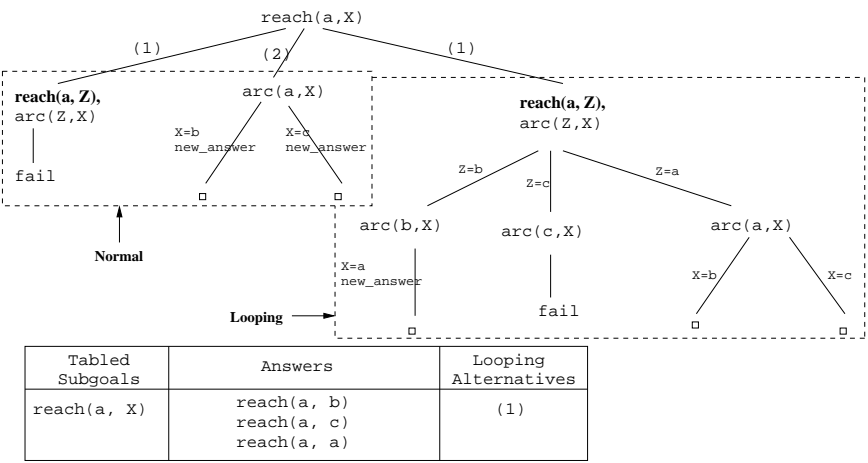,height=6cm,width=11.5cm}}
\caption{DRA Resolution for Example~\ref{ex:reach}(a)}
\label{fig:reach}
\end{figure}

As   shown    in Figure~\ref{fig:reach},   the    computation  of a tabled call
(e.g., {\tt reach(a,X)}) is divided into three  stages: {\it normal}, {\it looping}
and {\it complete}. The purpose of the normal stage is to find all the
looping alternatives (clause (1) leading to a variant subgoal {\tt
reach(a,Z)}) and record all the answers generated from the non-looping
alternatives  (clause (2)) into  the table.  The {\tt new\_answer}
label indicates   that the new  answer  generated from that successful
path should be added into the table. Then,  in the looping stage only
the looping alternative    (clause (1))  is  applied repeatedly  to
consume new tabled answers until a fixed-point is reached, that is, no
more answers  for  {\tt  reach(a,X)} can   be found.  Afterwards,  the
complete stage is reached. As a result, the  query {\tt :- reach(a,X)}
returns a complete  answer  set {\tt X=b}, {\tt   X=c} and {\tt  X=a},
albeit the predicate is defined left-recursively.

The DRA tabled resolution adopts  a dynamic  computation strategy by keeping
track of the type of the current resolvent. If the current resolvent 
is a non-tabled call (e.g., all the calls to the predicate {\tt arc/2}), 
then the traditional SLD resolution is applied.
Otherwise, if it is a tabled call, then  
the first occurrence of a tabled call in the computation is 
referred to as the {\em master} call; subsequent variant calls are called {\em slave} 
calls in DRA resolution. The master tabled call (e.g., {\tt reach(a,X)}) 
is responsible for exploring the matched clauses, manipulating execution states, 
and repeatedly applying its corresponding looping alternatives in
order to collect the tabled answers, whereas the slave tabled call 
(e.g., {\tt reach(a,Z)}) only consumes tabled answers if there are any in the table.

In DRA resolution, the procedure of computing fixed-points 
of a definite logic program mimics the semi-naive bottom-up 
computation strategy \cite{semi-naive},
whenever a looping alternative is applied again, the variant tabled 
calls only consume the incremental part of solutions in the table. 

\begin{Ex}
Consider the following tabled Prolog program with multiple looping alternatives:   
\label{ex:multi}
\end{Ex}
\begin{quote}
\begin{verbatim}
:- table r/2.
r(X, Y) :- r(X, Z), p(Z, Y).         (1)
r(X, Y) :- p(X, Y).                  (2)
r(X, Y) :- r(X, Z), q(Z, Y).         (3)
p(a, b).     p(b, c).     q(c, d).
:- r(a, Y).
\end{verbatim}
\end{quote}

\begin{figure}[htb]
\centerline{
\psfig{figure=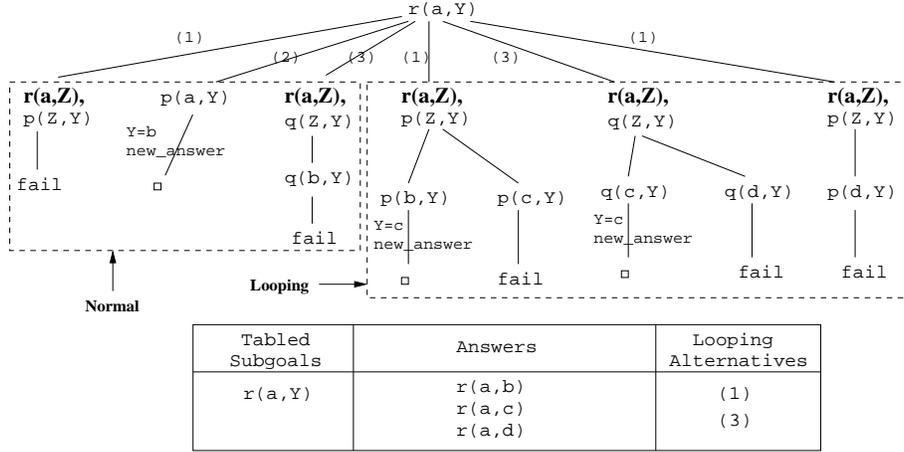,height=6cm,width=12cm}}
\caption{DRA Resolution for Example~\ref{ex:multi}}
\label{fig:multi}
\end{figure}

Figure~\ref{fig:multi} gives the  computation tree produced with
the DRA scheme for example~\ref{ex:multi}  (note that the  labels on the
branch refer to the clause  used  for creating that branch).  Both
clause (1)  and clause (3) need  to be tabled as  looping
alternatives for the  tabled call  {\tt   r(a,Y)}. The second alternative 
is   a non-looping alternative that  produces an answer for the   call {\tt r(a,Y)}
which is recorded in the table (denoted by the operation {\tt new\_answer}
in the figure). The query call {\tt r(a,Y)}  is a master tabled call  (since it is the
first occurrence), while all the occurrences of  {\tt r(a,Z)} are
slave tabled calls (since they are calls to  variant of {\tt
r(a,Y)}).  When the call {\tt r(a,Y)}  enters its looping state,
it keeps applying  the looping alternatives repeatedly until the
answer set does not change any more, i.e., until {\tt r(a,Y)}  is
completely evaluated. Note that if we added two more facts: {\tt
p(d,e)}  and {\tt q(e,f)}, then we will have  to go through  the two
looping  alternatives one  more  time to produce the answers {\tt
r(a,e)} and {\tt r(a,f)}. Each time a looping alternative is invoked,
only incremental tabled solutions are consumed.

\subsection{A Semi-naive Algorithm for DRA}

We next give a meta-interpreter to describe how DRA works operationally.
To  keep  the presentation and the meta-interpreter simple and  clear,  
we assume  that there  are no nested tabled calls dependent  on  each other
and we ignore any further optimizations except the incremental consumption for
the semi-naive algorithm. We also assume that the table exists as a global data structure.    

As shown in Figure~\ref{fig:metai}, the {\tt  solve/1}  predicate  is  the entry point
to    the meta-interpreter   and takes as input the goal to be executed.
The  goal {\tt clause(A,Cl)}  nondeterministically  finds  the
matching clause,  {\tt Cl},  for   the    goal  {\tt A}, and  
then {\tt setCurrentAlt(A,Cl)} records
the current alternative {\tt Cl} associated with {\tt A} in a temporary 
global database. 
The  goal {\tt tabled(A)} checks whether {\tt A} has been declared as tabled or not,
{\tt state(A,X)} checks to  see  whether {\tt  A}'s  execution
status is {\tt  X} (one of {\tt normal}, {\tt   looping},   or   {\tt
complete}), while the goal {\tt setState(A,X)} changes the execution
state of the goal {\tt A} to the state {\tt X}.  The goal {\tt
addTableAns(A)} adds an answer for goal {\tt A} to the table (if
it is not already present), while the goal {\tt look\_up\_table(A)}
looks up answers for {\tt A} that have been
recorded so  far in the table. The goal {\tt addLoopAlt(A,Cl)} finds the 
current alternative of {\tt A} (which was previously recorded by {\tt setCurrentAlt(A,Cl)}), 
and then tables it as  a looping alternative for {\tt  A}.  
The  goal   {\tt getLoopAlt(A,La)} nondeterministically  finds the next looping
alternative, {\tt La}, from the list of {\tt A}'s looping alternatives.

\begin{figure*}[tb!]
{{\small\tt
(1)~~solve(true).\\
(2)~~solve((A,B)) :- solve(A),\ solve(B).\\
(3)~~solve(A) :-\\
(4)~~~~~~~(tabled(A) ->\\
(5)~~~~~~~~~~~(state(A, normal) -> ~~~~~~~~~\%\%\ {\tt A} in {\em normal} state\\
(6)~~~~~~~~~~~~~~~(type(A, master) -> ~~~~~~\%\% {\tt A} is a {\em master} call\\
(7)~~~~~~~~~~~~~~~~~~~(clause(A,Cl) ->\\
(8)~~~~~~~~~~~~~~~~~~~~~~~setCurrentAlt(A,Cl), solve(Cl), addTableAns(A)\\
(9)~~~~~~~~~~~~~~~~~~~;~~~setState(A, looping), solve(A)\\
(10)~~~~~~~~~~~~~~~~~~)\\
(11)~~~~~~~~~~~~~~;~addLoopAlt(A), ~~~~~~~~~\%\% {\tt A} is a {\em slave} call\\
(12)~~~~~~~~~~~~~~~~look\_up\_table(A)\\
(13)~~~~~~~~~~~~~~)\\
(14)~~~~~~~~~~;~(state(A, looping) -> ~~~~~~\%\% {\tt A} in {\em looping} state\\
(15)~~~~~~~~~~~~~~(type(A, master) -> ~~~~~~\%\% {\tt A} is a {\em master} call\\
(16)~~~~~~~~~~~~~~~~~~(getLoopAlt(A, La) ->\\
(17)~~~~~~~~~~~~~~~~~~~~~~solve(La), addTableAns(A)\\
(18)~~~~~~~~~~~~~~~~~~;~~~(isNewSolFnd(A) ->\\
(19)~~~~~~~~~~~~~~~~~~~~~~~~~~solve(A)\\
(20)~~~~~~~~~~~~~~~~~~~~~~;~~~setState(A, complete)\\
(21)~~~~~~~~~~~~~~~~~~~~~~)\\
(22)~~~~~~~~~~~~~~~~~~)\\
(23)~~~~~~~~~~~~~~;~~~look\_up\_table(A) ~~~~~~\%\% {\tt A} is a {\em slave} call\\
(24)~~~~~~~~~~~~~~)\\
(25)~~~~~~~~~~;~look\_up\_table(A) ~~~~~~~~~~~~\%\% {\tt A} in {\em complete} state\\
(26)~~~~~~~~~~))\\
(27)~~~~~~; clause(A, Cl), solve(Cl)~~~~~~~\,\,\,\,\%\% non-tabled call\\
(28)~~~~~~).
}}
\caption{An algorithm for DRA resolution}
\label{fig:metai}
\end{figure*}

The meta-interpreter  in Figure~\ref{fig:metai} is
pretty self-explanatory.  A non-tabled goal is computed as in standard
Prolog, with leftmost-first selection rule (line  2) and a depth-first
search rule (line 27).
A  tabled call, on the other hand, is resolved in a different manner  by the
DRA scheme based
on its  execution  state (normal, looping  or  complete)  and its {\em
master-slave}  mode. Consider a tabled goal  {\tt A} (line 4).  If {\tt
A}'s execution  state is normal   (line 5) and it is a  master  call
(line  6), then a matching clause {\tt  Cl} for {\tt  A} is
nondeterministically found, and the subgoals in the  
body of {\tt Cl} recursively solved (line 7-8).
An answer for {\tt A} will be added into the
table  if its matching clause {\tt Cl} is successfully solved (line 8). The
matching clause {\tt Cl} is identified as a  looping alternative
if a slave  call of {\tt  A}  is found (line 11), and the  slave
call can only be  resolved against tabled  answers (line 12).
After  all matching  clauses   have been seen, {\tt A}'s execution
state is set  to {\tt looping} (line  9). During the looping state
(line 14), the master call of {\tt A} retrieves and executes the
collected looping alternatives (line 15-17); whereas the slave
call of {\tt A} can only consume the tabled answers. If no new
answer is found while  applying one cycle of all the  looping
alternatives w.r.t.  the master call of   {\tt A}, the   execution
state is  set to {\tt complete} (line 20); otherwise goal {\tt A}
continues to be executed (line 19) during its looping state until
no  new answer can  be found. Once the  tabled goal {\tt A}
reaches  its complete state,  all answers would have been recorded in the
table, and thus   only  table look-ups  are   used for
resolving subsequent variant calls to A.

To make the algorithm efficient, we introduce an incremental consumption
strategy for the predicate {\tt look\_up\_table/1} to mimic semi-naive 
evaluation of logic programs \cite{semi-naive}.
Incremental consumption refers to avoiding re-consuming older tabled answers 
which have already been consumed earlier by the same slave tabled calls.
Each looping alternative maintains two pointers, called {\tt begin} and {\tt end},
which divide the tabled answer list into three parts: old answers, 
current answers, and new answers. Old answers are those tabled answers that 
were consumed during the previous round of exploring the looping alternative;
current answers can be consumed during the current exploration; and
new answers are kept for the next round of consumption.
The answers between the pointers {\tt begin} and {\tt end} constitute
the incremental answer set to be consumed in the current round. 
The incremental answer set is updated dynamically whenever a looping alternative 
is picked up for exploration. 
Therefore, the predicate {\tt look\_up\_table/1} only consumes the
incremental answer set.

\subsection{Experiments}

The DRA resolution builds the computation tree as in normal Prolog execution 
based on SLD, however, when a variant tabled call is encountered, 
the branch that leads to that variant call is tabled as a looping alternative,
and later applied again during the looping state. This has the same effect as
shifting branches with variant calls to the right of the SLD tree to avoid
exploring potentially non-terminating branches. 
Because of this simplicity, the DRA resolution can be incorporated
very easily and without sacrificing efficiency in an existing Prolog system. 
This can have important consequences, given that tabling is so important for 
many serious applications of logic programming (e.g., model checking \cite{cav-iv}). 
The simplicity of our scheme guarantees that execution is
not inordinately slowed down (e.g., in the previous B-Prolog tabled system \cite{zhou}, 
a tabled call may have to be re-executed several times to ensure that 
all answers are found), nor considerable amount of memory
used (e.g., in the XSB tabled system \cite{xsb1} a large number of stacks/heaps 
may be frozen at any given time), rather, the raw speed of the Prolog's 
WAM engine is available to execute even those programs that contain variant calls.
The new tabling scheme allows one to incorporate tabling in an existing 
logic programming system with very little effort. Using the DRA scheme we were 
able to incorporate tabling in the commercial ALS Prolog system \cite{als} 
in a few man-months of work. The time efficiency of our tabled ALS (called TALS)
system is comparable to that of the XSB system. The space efficiency of our system
is better than that of XSB system. 

\begin{table}[htb]
\begin{center}
\begin{tabular}{|c|c|c|}
\hline\hline
{\bf Benchmarks} &  \emph{TALS} & \emph{XSB}\\
\hline\hline
{\bf cs\_r} & 0.37/29.8 & 0.20/58.3 \\
\hline
{\bf disj} &  0.25/18.7 &  0.05/54.2 \\
\hline
{\bf kalah}	&  0.20/37.0 &  0.05/97.1 \\
\hline
{\bf peep} & 0.47/24.0 &  0.18/376.3 \\
\hline
{\bf pg} & 0.29/23.9 &  0.05/150.5 \\
\hline
{\bf read} & 0.62/35.8 & 0.23/616.7 \\
\hline
{\bf sg} & 0.03/27.4 & 0.08/48.3 \\
\hline\hline
\end{tabular}
\caption{Running-time(Seconds)/Space-usage(KB)}
\label{tbl:compare}
\end{center}
\end{table}

A preliminary implementation of DRA in the ALS Prolog system has been completed. 
Table~\ref{tbl:compare} gives the comparison of running times (in seconds)
between XSB and TALS for various benchmarks. These benchmarks are
distributed with XSB and most of them table multiple predicates,
many of whom manipulate structures. In general, the time performance of TALS 
is worse than that of XSB. That is mainly because SLG resolution 
has  been implemented  by a combination of computation suspension 
via  {\it  stack  freezing} and maintaining  a forest of SLD trees. 
Due to the computation suspensions and resumption, XSB avoids 
reconstructing execution environment for applying looping alternatives,
which is typically recomputed in TALS since looping alternatives
have to be applied again to ensure the completion of fixed-point computation.
On the other hand, due to this maintenance of forest of SLD trees and 
freezing of  stacks, XSB cannot be implemented   in  the  same  way as 
SLD resolution  using  a simple stack-based memory structure. Consequently, 
the freezing of stacks results in significant space overheads.
It should also be mentioned that DRA has been implemented
on top of ALS system without modifying the compiler. 
XSB includes tabling in compiling stage, which may have 
significant impact on performance.

Tables~\ref{tbl:compare} also compares the space usage
between TALS and XSB systems. The space includes total stack and 
heap space used as well as {\it space overhead} to
support tabling. The space overhead to support tabling in case of TALS 
includes table space,  the extra space needed to record looping alternatives 
and extra fields for keeping track of the types of tabled calls. 
In case of XSB, the figure includes the table space and
space used for suspension in SLG-WAM \cite{SW94}. 
As can be noticed from Table~\ref{tbl:compare},
the space performance of TALS is significantly better than that
of XSB (for some benchmarks, e.g., peep, pg and read,
it is orders of magnitude better). 

\section{Flexible Fixed Point Computation via Modes}
\label{sec:mode}

There are many situations in which a fixed-point contains 
a large number or even infinite number of solutions 
(for example~\ref{ex:reach}(b)). In these cases, a fixed-point 
computation engine may not be efficient enough or feasible at all. 
In this section, we present a mode-declaration scheme \cite{mode} which provides
the users the capability to reduce a fixed-point from a large solution set to 
a preferred small one, or from an infeasible infinite set to a finite one.  
The mode declaration scheme can be characterized as a meta-level
mapping operation over the original fixed-point.

\subsection{Mode Declarations}

The fixed-point reduction can be achieved by a mode declaration for tabled
predicates, which is described in the form of\\
\centerline{\tt :- table $q(m_1, ...,  m_n)$.} \\
where $q/n$ is a tabled  predicate name, $n \geq  0$, and each $m_i$ has
one of the forms as defined in Table~\ref{tbl:modes}. 

\begin{table}[htb]
        \centering
  \begin{tabular}{|c|c|}
    \hline
    \hline
    Modes & Informal Semantics\\
    \hline
    \hline
    $\mathbf{+}$ & an indexed argument\\
    $\mathbf{-}$ & a non-indexed argument\\
    {\bf min} & a minimum non-indexed argument\\
    {\bf max} & a maximum non-indexed argument\\
    \hline
    \hline
  \end{tabular}
  \caption{Built-in Modes for Tabled Predicates}
  \label{tbl:modes}
\end{table}

In order to find out how modes will affect 
fixed-point computations, we have to better understand the
function of variant checking in tabled resolution. 
Variant  checking is a crucial  operation  for tabled resolution as it
leads  to avoidance of non-termination. It is  used to differentiate
between the various tabled calls as well as their answers.  
While computing the answers to a
tabled goal $p$ with tabled resolution,  if another tabled subgoal $q$
is  encountered, the decision  regarding   whether to consume   tabled
answers  or to try program clauses   depends on the  result of variant
checking.  If $q$ is a variant of $p$, the variant subgoal $q$ will be
resolved by unifying it  with  tabled answers, otherwise,  traditional
Prolog resolution is adopted for $q$. Additionally,  when an answer to
a tabled goal is generated, variant  checking is used to check whether
the generated answer is variant of  an answer that is already recorded
in the table. If so, the table is not changed; this step is crucial in
ensuring that a fixed-point is reached.

\begin{figure}[htb]
\centerline{
\psfig{figure=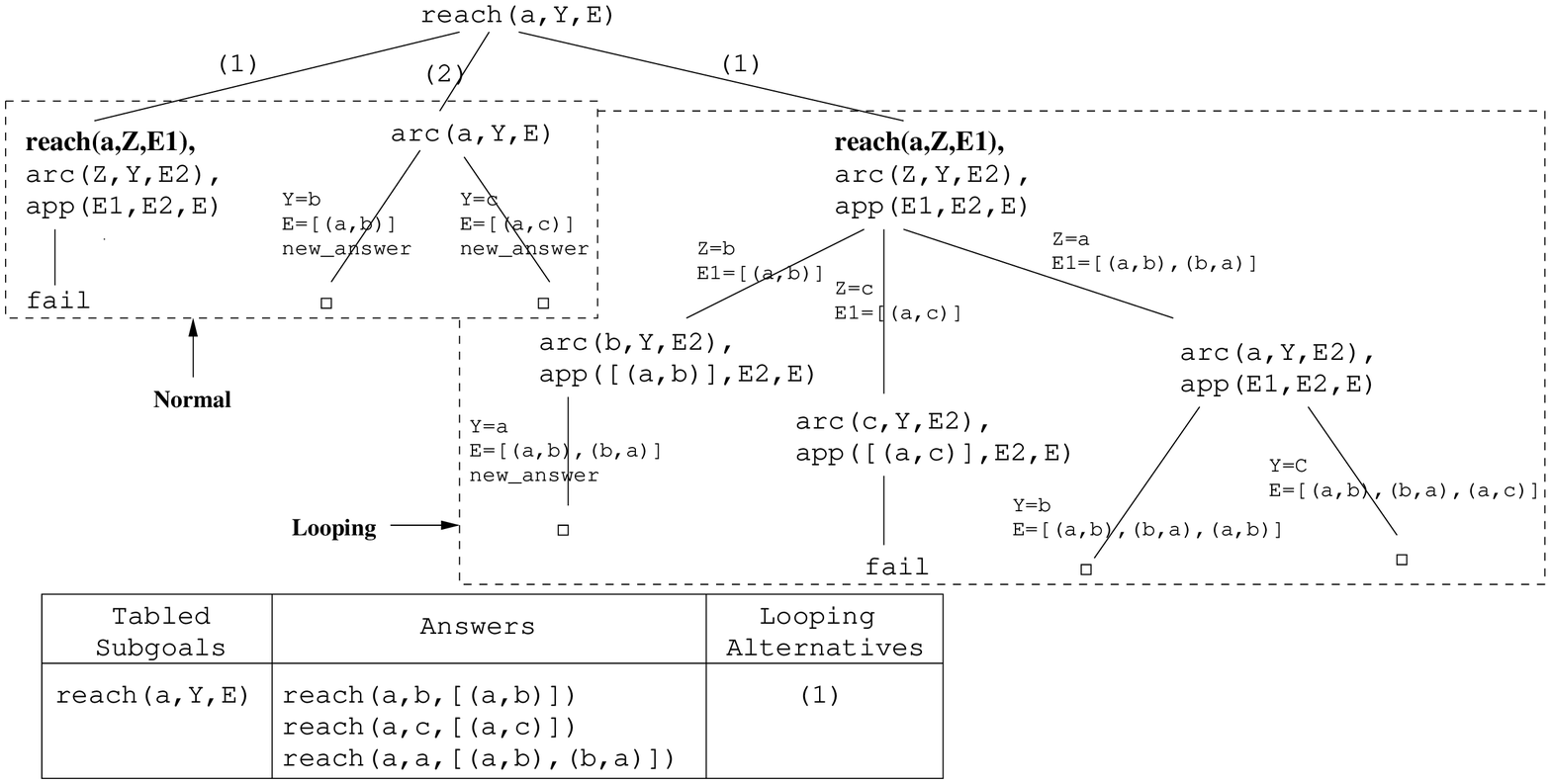,height=6.2cm,width=11.4cm}}
\caption{DRA Resolution with Mode Declaration}
\label{fig:example}
\end{figure}

The main purpose of mode declaration is to classify the predicate
arguments into two types: {\em indexed} and {\em non-indexed}.
Only indexed arguments are used for variant checking while collecting
answers for the table;  for each tabled call, any answer generated 
later for the same value of the indexed arguments is discarded 
because it is a variant, w.r.t. the indexed arguments, 
of a previously tabled answer.
Consider again the reachability program in Example~\ref{ex:reach}(b). 
Suppose we declare the mode as ``{\tt :- table reach(+,+,-)}''; this means
that only the first two arguments of the predicate {\tt reach/3} are used
for variant checking. The new  computation of the query {\tt reach(a,Y,E)}
is shown  in  Figure~\ref{fig:example}.   Since  only the   first  two
arguments of {\tt reach/3} are used for variant checking, the last two
answers    ``{\tt  Y=b,   E=[(a,b),(b,a),(a,b)]}''   and  ``{\tt  Y=c,
E=[(a,b),(b,a),(a,c)]}'', shown on the rightmost two sub-branches, are
variant   answers  to  ``{\tt   Y=b,    E=[(a,b)]}'' and  ``{\tt  Y=c,
E=[(a,c)]}'' respectively.  Therefore,  no new answers are added  into
the table at those points. The computation is then terminated properly
with three answers.  As a result, each reachable node from {\tt a} has
a simple path as an explanation.

The mode directive {\tt table} makes it very easy and efficient
to extract explanation for tabled predicates. In fact, our strategy of
ignoring the explanation  argument during variant  checking results in
only the    first  explanation  for  each  tabled     answer  being
recorded. Subsequent explanations are filtered by our modified variant
checking    scheme.  This   feature     ensures that those   generated
explanations are concise  and that cyclic explanations  are guaranteed
to be absent. For the  reachability instance shown in 
Figure~\ref{fig:example}, each returned path is simple so 
that all arcs are distinct.

Essentially, if  we regard a tabled   predicate as a function, then  
all the non-indexed arguments are uniquely defined  by the instances of 
indexed arguments. For the previous example, the third argument of 
{\tt reach/3}  returns a single path depending on the first two arguments.   
Therefore, variant checking should   be  done w.r.t. only indexed
arguments during tabled resolution. From this viewpoint, the mode
declaration makes tabled resolution more efficient and flexible.
More importantly, this declaration scheme is especially useful
in reducing an infinite set of solutions to a finite one for some
practical uses, or to reduce a large finite set of solutions to
an optimized smaller set as shown next.

\subsection{Declaration of Aggregates}

The mode directive   {\tt table}  can  be  further extended  to
associate a  non-indexed argument of   a  tabled predicate  with  some
aggregate constraint. With the mode `{\tt -}',  a non-indexed argument for each tabled
answer  only records  the very  first instance.  This  ``very first''
property can actually be generalized to support other preferences, 
e.g., the minimum value with mode {\tt min} (or the maximum with mode {\tt max}), 
in  which case the  global table will  record  answers with the value 
of that argument as small (or great) as possible. 
That is, a tabled answer can be dynamically replaced by a new one with 
smaller (or greater) value during the computation. 
Note that we only enumerate two typical aggregates as examples, other aggregates,
such as {\em sum}, {\em average}, or even a user-defined one, can be used as well. 

The aggregates, {\tt min} and {\tt max}, are specified via mode declarations 
as shown in Table~\ref{tbl:modes}). Both modes imply that the declared arguments 
are non-indexed. The aggregates can be used to specify
optimization problems more elegantly.
For instance, in the code shown in example ~\ref{ex:short} for
searching for shortest paths, instead of defining
the shortest path directly, we only need to specify what is the definition for  
a general path. 
Clauses (2) to (5)  make up the core program defining the path relation
and a directed graph with a set of edges; 
Clause (1) specifies the predicate {\tt path/4} to be optimized and 
gives the criteria regarding how it should be optimized.
The mode declaration {\tt path(+,+,min,-)} means that only the first two
arguments (pair of nodes) are used for variant checking when an answer
is generated, and a minimum value (the shortest path) is expected for 
the third argument. Arguments  with  different modes   are  tested 
in the  following order during  variant  checking  of a  recently  
generated answer:  (i)  the indexed argument  with `{\tt +}' mode has  
the  highest priority to be
checked to identify whether it is a  new answer. If that is the
case, a new tabled entry is required to record the answer; otherwise a
tabled answer with the same indexed arguments is found. 
(ii) This tabled answer   is then  compared   with the  recently  
generated  one w.r.t the argument  with  the optimum mode  `{\tt min}';  
if the  new answer has a smaller value for this argument,  
then a  replacement of the tabled answer is required such that 
the tabled  answer keeps the minimum value as expected for 
this argument.
(iii) The last argument with mode `{\tt -}' will not be used for variant 
checking; if a replacement of a tabled answer happens, then the argument
will be replaced as well; otherwise, the recently generated answer as well
as its fourth argument are discarded.

\begin{Ex}
\label{ex:short}
Consider the  following program searching for a shortest path, where
{\tt path(X,Y,D,L)} denotes a path from {\tt X} to {\tt Y} with the distance {\tt D}
and the path route {\tt L}.
\end{Ex}

\begin{quote}
\begin{verbatim}
:- table path(+, +, min, -).                        (1)
path(X, X, 0, []).                                  (2)
path(X, Y, D, [e(X, Y)]) :-  edge(X, Y, D).         (3)
path(X, Y, D, [e(X, Z) | P]) :-
     edge(X, Z, D1), path(Z, Y, D2, P), 
     D is D1 + D2.                                  (4)
edge(a,b,4).    edge(b,a,3).    edge(b,c,2).        (5)
:- path(a, X, D, P).                                (6)
\end{verbatim}
\end{quote}

Example~\ref{ex:short} shows that even though the core program
defines a general path, with mode declaration, the predicate {\tt path/4}
can be easily upgraded to an optimization predicate.
As long as the tabled Prolog engine is set to  compute the fixed-point
semantics  for logic  programs, the shortest path under consideration will
always be  found. Intuitively, given a  tabled call $\mathcal{C}$, the
DRA resolution first finds   all the answers for   $\mathcal{C}$ using
clauses  not containing  variant calls.  Once  this set  of answers is
computed and tabled,  it is treated as  a set of  facts, and used  for
computing rest of the  answers  from the  clauses leading  to  variant
calls  (looping alternatives). Whenever an  answer to $\mathcal{C}$ is
generated, it  will be selectively added to  the table either as a new
entry  or   as a replacement    based   on the defined  mode   of  the
corresponding predicate.  The process stops when no new answers can be
computed  via the   looping  alternatives,  i.e.,  a  fixed  point  is
reached. In  this regard, with the  assistance of mode declaration and
tabled resolution,  the computation  of program clauses  only defining
general solutions will still produce the optimal solution.

\subsection{Operational Semantics}

The operational semantics of a tabled program is dependent on 
tabled resolution \cite{chen,zhou,tals}, 
which can be formalized based on the Herbrand model 
\cite{van,lloyd} and fixed-point theory. 
 In spite of having different tabled resolution, a tabled Prolog can be thought of 
as an engine for efficiently computing the least fixed-points.
The procedure of computing fixed-points of a definite logic program 
in DRA resolution mimics the bottom-up computation strategy as follows.
For the consideration of clarity and simplicity, we ignore any optimization
used in DRA resolution (e.g. incremental consumption for simulating semi-naive bottom-up computation).

We use the following notational conventions: $P$ is used to denote a tabled
logic program, $B_P$ to denote the Herbrand base of $P$, $2^{B^P}$ to denote
the set of all Herbrand interpretations of $P$, a ground instance (e.g., a ground atom,
a ground instance of a clause) to denote an instance without involving 
any variable. Note that $\omega$ is the first infinite ordinal, 
and $(\mathcal{F}\uparrow n)(x)$ to denote
applying the mapping $\mathcal{F}$ $n$ times as
$\overbrace{\mathcal{F}(\mathcal{F}(\cdots \mathcal{F}}^n(x)\cdots))$.

\begin{mydef}
Let $P$ be a logic program and $B_P$ its Herbrand base. Let $\emptyset$ denote the
empty set.
We define a meta-level procedure $T_P:2^{B_P} \rightarrow 2^{B_P}$. 
Given a Herbrand interpretation $I$,  $T_P(I)$ performs:\\
\verb~   ~ 1. $I_0 \leftarrow \emptyset;$\\
\verb~   ~ 2. {\bf for} each ground instance $A\ \mbox{:-}\ A_1, \cdots, A_m$ of a clause in $P$ \\
\verb~      ~where $\{A_1,\cdots,A_n\} \subseteq I$, {\bf do}\\
\verb~       ~ $I_0 \leftarrow I_0 \cup \{A\}$;\\
\verb~   ~ 3. {\bf return} $I_0$.\\
The fixed-point semantics of $P$ can be described as 
$T_P\uparrow\omega(\emptyset)$ {\tt \cite{lloyd}}.
\label{def:fixed}
\end{mydef}

We next show that how mode declaration affects the 
fixed-point semantics of a logic program. One key ingredient for 
the mode declaration scheme to be applicable is the {\em optimal-substructure} property
\footnote{A problem has optimal-substructure property if its optimal
solution can be expressed in terms of optimal solutions of its subproblems.},
that is, the optimal solution to a tabled call contains optimal solutions
to its tabled sub-calls. Typical examples of such problems are those for dynamic programming.
For simplicity, we assume that for any tabled predicate, there is at most one optimization mode, 
`{\tt min}' or `{\tt max}', in the mode declaration. Tabled predicates whose multiple arguments
have an optimization mode can be handled by combining them into one 
via program clause transformation. 
Additionally, we assign non-indexed modes different priorities, i.e., `{\tt min}'
and `{\tt max}' have higher priorities than `{\tt -}'. 

Note that a tabled predicate without explicit mode declaration has 
a default one with all indexed modes `{\tt +}'. 
Although mode declarations are only allowed for tabled predicates, 
non-tabled predicates can also be simply treated as having implicitly 
declared indexed mode `{\tt +}' for all arguments. Thus, 
in the rest of this subsection we will not distinguish tabled predicates
from non-tabled ones, and each predicate defined in a tabled 
logic program is associated with a mode declaration.

\begin{mydef}
Let $q/n$ be a predicate with a mode declaration $q(m_1, m_2, \cdots, m_n)$ 
in a tabled logic program $P$. Let the arguments 
$m_{i1}, m_{i2}, \cdots, m_{ik}\ (0 \leq k \leq n)$ 
have the mode `{\tt +}' such that $1 \leq i1 < i2 < \cdots < ik \leq n$;
let $m_j$ be the argument with the highest priority non-indexed mode if there 
are non-zero arguments in $q/n$ with non-indexed modes. 
We define two functions $\mathcal{K}_{q/n}$ and $\mathcal{O}_{q/n}$ as follows:
given a ground atom $q(a_1, a_2, \cdots, a_n)$, 
\begin{eqnarray*}
\mathcal{K}_{q/n}(q(a_1, a_2, \cdots, a_n)) & = & (a_{i1}, a_{i2}, \cdots, a_{ik});\\
\mathcal{O}_{q/n}(q(a_1, a_2, \cdots, a_n)) & = & a_{j},\ \ \ \mbox{if\ \ $m_j$ exists.} 
\end{eqnarray*}
We say two ground atoms, $t_1$ and $t_2$, of $q/n$ {\em comparable} if and only if 
$\mathcal{K}_{q/n}(t_1) = \mathcal{K}_{q/n}(t_2)$.
\end{mydef}

The function $\mathcal{K}_{q/n}$ is used to return a sequence of indexed arguments in a 
left-to-right order, whereas $\mathcal{O}_{q/n}$ returns a non-indexed argument with the highest
priority. We say two ground atoms of $q/n$ are {\em comparable} if and only if 
these two atoms have the same indexed arguments.
We abbreviate $\mathcal{K}_{q/n}$ and $\mathcal{O}_{q/n}$ 
to $\mathcal{K}$ and $\mathcal{O}$, respectively, 
whenever the predicate is obvious from the context.
Thus, we have a preference relation defined as follows. 

\begin{mydef} 
Let $P$ be a logic program. 
A {\em preference relation} in $P$ is a strict partial order
\footnote{A strict partial order relation is irreflexive and transitive.}
relation $\prec_{P}$ s.t. for any two ground atoms $A_1$ and $A_2$ of a 
 predicate $q/n$ in $P$, $A_1 \prec_{P} A_2$ if both of the followings are true:
\begin{itemize}
\item $\mathcal{K}(A_1) = \mathcal{K}(A_2)$;
\item {\bf cases} the non-indexed mode with the highest priority in $q/n$ {\bf of} \\
     {\em min:}  $\mathcal{O}(A_1) > \mathcal{O}(A_2)$;\\
     {\em max:}  $\mathcal{O}(A_1) < \mathcal{O}(A_2)$;\\
     {\em $-$:}    $A_2$ is generated earlier than $A_1$ during tabled resolution.
\end{itemize}
\label{def:prefer}
\end{mydef}

We abbreviate $\prec_{P}$ to $\prec$ whenever the tabled logic program 
is obvious from the context. It should be mentioned that the semantics 
of mode `{\tt $-$}' is heavily dependent on the order in which the answers 
are generated, which is decided by the tabled resolution procedure used. 
For Example~\ref{ex:short}, the preference relation $\prec$ 
is the set 

{
\[\begin{array}{l}
\{ ~\mbox{path}(a,a,7,\_) \prec \mbox{path}(a,a,0,\_), \\
   \mbox{path}(a,a,14,\_) \prec \mbox{path}(a,a,0,\_), ~ \mbox{path}(a,a,14,\_)
\prec \mbox{path}(a,a,7,\_),\\         ~~~ \cdots\\    
\mbox{path}(a,b,11,\_) \prec \mbox{path}(a,b,4,\_), \\
\mbox{path}(a,b,18,\_) \prec \mbox{path}(a,b,4,\_), ~ \mbox{path}(a,b,18,\_) \prec
\mbox{path}(a,b,11,\_),\\
~~~ \cdots ~ \cdots \}
\end{array}
 \]
}\\
where the numbers $0, 7, 11, ...$ are the possible distances for their corresponding
pair of nodes, and `$\_$' means any ground term from the Herbrand universe. 
Note that no atoms of the non-tabled predicate {\tt edge/3} are in the preference relation.
This is because each non-tabled predicate is implicitly declared to have the index mode
`{\tt +}' for all its arguments; therefore, none such predicate can satisfy the second condition 
of Definition~\ref{def:prefer}. Similarly, all ground atoms of non-tabled predicates
are {\em optimized} according to the following definition.

\begin{mydef}
Let $P$ be a logic program and $I$ be one of its Herbrand interpretations;
We say that $A$ is {\em an optimized ground atom}, abbreviated as {\em an optimized atom}, 
in $I$ if there does not exist any other ground atom $A_1 \in I$ s.t. $A \prec A_1$.
\label{def:optimized} 
\end{mydef}

\begin{mydef}
Let $P$ and $B_P$ be a logic program and its Herbrand base. We define a meta-level procedure 
$T_P^{'}:2^{B_P} \rightarrow 2^{B_P}$. Given a Herbrand interpretation $I$, 
$T_P^{'}(I)$ performs:\\
\verb~   ~ 1. $I_0 \leftarrow \emptyset;$\\
\verb~   ~ 2. {\bf for} each ground instance $A\ \mbox{:-}\ A_1, \cdots, A_m$ of a clause in $P$ \\
\verb~      ~where $\{A_1,\cdots,A_n\} \subseteq I$, {\bf do}\\
\verb~      ~ 2a. $I_0 \leftarrow I_0 \cup \{A\}$;\\
\verb~      ~ 2b. $I_0 \leftarrow I_0 - \{a_1 \in I_0 : \exists a_2 \in I_0 \mbox{ s.t. } a_1 \prec a_2\}^{\ (*)}$\\
\verb~   ~ 3. {\bf return} $I_0$.\\
Thus, the fixed-point semantics of $P$ can be described as 
$T_P^{'}\uparrow\omega(\emptyset)$.
\label{def:newfixed}
\end{mydef}

Def.~\ref{def:newfixed} gives the fixed-point semantics for a tabled logic program
with mode declaration. The statement (*) shows how non-indexed modes affect the 
the procedural semantics of the core program through the preference relation $\prec$.

\noindent
\begin{pro}
Let $P$ be a tabled logic program. For any atom $A \in T_{P}^{'}\uparrow n(\emptyset)$,
where $n \geq 0$, $A$ is an optimized atom in $T_{P}^{'}\uparrow n(\emptyset)$.
\label{pro:newoptimized}
\end{pro}

\noindent {\bf Proof}:\ \ 
This can be easily shown by mathematical induction on $n$, mainly 
using the result of step $2b$ in Definition~\ref{def:newfixed}:\\
\centerline{$I_0 \leftarrow I_0 - \{a_1 \in I_0 : \exists a_2 \in I_0 \mbox{ s.t. } a_1 \prec a_2\}$,}\\
so that any atom $A$ in the resulting $I_0$ is an optimized atom according to Definition~\ref{def:optimized}.
\hfill $\Box$

\noindent
\begin{pro}
Let $P$ be a tabled logic program. If $A$ is an optimized atom in 
$T_P\uparrow n(\emptyset)$, then we have $A \in T_{P}^{'}\uparrow n(\emptyset)$, 
for any $n \geq 0$.
\label{pro:correct}
\end{pro}

\noindent {\bf Proof}:\ \ 
The proof is based on a mathematical induction on $n$.\\
{\tt Base case:} Consider $n=0$. Since $T_P\uparrow 0(\emptyset)$ 
is an empty set, the proposition is vacuously true. \\
{\tt Inductive Case:} Assume that the proposition is true for some $i \geq 0$. 
We consider an optimized atom $A \in T_P\uparrow\ (i+1)(\emptyset)$. 
$A$ is obviously an optimized atom in $T_P^{'}\uparrow (i+1)(\emptyset)$ as well 
due to the fact that $T_P^{'}\uparrow (i+1)(\emptyset) \subseteq T_P\uparrow\ (i+1)(\emptyset)$.
Next, we complete the proof by showing $A \in T_P^{'}\uparrow (i+1)(\emptyset)$. 
According to Definition~\ref{def:fixed}, there exists a ground instance
$A \mbox{:-} A_1, ..., A_m$ (for some $m \geq 0$) of a clause in $P$ where
$\{A_1, ..., A_m\} \subseteq T_P\uparrow i(\emptyset)$. Based on the
optimal-substructure property, $A_1, ..., A_m$ must be optimized atoms
in $T_P\uparrow i(\emptyset)$. Following the induction assumption, 
we have $\{A_1, ..., A_m\} \in T_{P}^{'}\uparrow i(\emptyset)$. 
Therefore, $A$ satisfies the conditions specified in Definition~\ref{def:newfixed};  
we have $A \in T_P^{'}\uparrow (i+1)(\emptyset)$. \hfill $\Box$

\noindent
\begin{pro}
Let $P$ be a tabled logic program. If $A \in T_{P}^{'}\uparrow n(\emptyset)$, 
then we have that $A$ is an optimized atom in $T_{P}\uparrow n(\emptyset)$, for any $n \geq 0$.
\label{pro:correct1}
\end{pro}

\noindent {\bf Proof}:\ \ 
We can easily get $A \in T_{P}\uparrow n(\emptyset)$ due to the fact that
$T_{P}^{'}\uparrow n(\emptyset)$ is a subset of $T_{P}\uparrow n(\emptyset)$. 
Assume that $A'$ is an optimized atom in $T_{P}\uparrow n(\emptyset)$ and 
$A \prec A'$. Based on Proposition~\ref{pro:correct}, we have 
$A' \in T_{P}^{'}\uparrow n(\emptyset)$, which is a contradiction with 
the fact that $A \in T_P^{'}\uparrow n(\emptyset)$ since  $A \prec A'$.
Therefore, $A$ must be an optimized atom in $T_{P}\uparrow n(\emptyset)$.
 \hfill $\Box$

Thus, we have the following main result
showing the correctness of our mode declaration scheme for solving
optimization problems with optimal-substructure property.

\begin{theorem}
Let $P$ be a tabled logic program. $A \in T_{P}^{'}\uparrow\omega(\emptyset)$
if and only if $A$ is an optimized atom in $T_{P}\uparrow\omega(\emptyset)$.
\label{thm:correct}
\end{theorem}

\noindent {\bf Proof}:\ \ 
The proof follows trivially from 
Proposition~\ref{pro:correct}  and Proposition~\ref{pro:correct1}.
 \hfill $\Box$

\subsection{Implementation}
\label{sec:implementation}

The mode  declaration  scheme has been implemented   in the authors'
TALS system, a  tabled Prolog system  implemented on  
the top  of the WAM engine of the commercial ALS Prolog engine \cite{als}.  
No change is required to the DRA resolution mechanism; 
therefore, the same idea can also be applied to other tabled Prolog systems such
as XSB and B-Prolog.

In the TALS system, the global data structure {\em table} is 
efficiently implemented thanks to the {\it trie} data 
structure \cite{tries}. The organization of the table can be 
abstractly described as a two-level hierarchical structure.
The first level is used to organize different tabled calls indexed by
their corresponding predicates names and arities; whereas the
second level is used to organize tabled answers indexed by 
the tabled calls. Variant checking is the main operation used to
locate the correct table position to access the table. 

Two major   changes  to  the global data  structure {\em  table} are
needed to support mode declarations. First, each table predicate
is associated with  a new item {\em mode},  which is represented as a
bit string. The default mode for each argument in a table predicate is
`{\tt +}'.  Second, the  answers  to a  tabled call are selectively
recorded depending on its mode declaration. 
The declared modes essentially specify
the user preferences or selection constraints among the answers. 
When a new answer to a tabled goal is generated, variant checking on indexed
arguments is invoked to determine whether the answer is variant to a
previously tabled one. If that is the case, declared modes on non-indexed arguments
are used to select a better answer to table; otherwise, a new table entry
is added to record the answer. In fact, if an indexed argument is
instantiated in advance before a tabled goal is called, variant checking on
this indexed argument can be avoided since its value is same for all the answers;
furthermore, it is not necessary to record the pre-instantiated value 
with each tabled answer because the same value has already been stored 
in the tabled call entry. This optimization leads to improvements on 
both time and space system performance.

Another important implementation issue is the replacement of 
tabled answers. In the current TALS system, if the tabled subgoal only
involves numerals as arguments, then the tabled answer will be completely
replaced if necessary. If the arguments involve structures, however,
then the answer will be updated by a link to the new answer. Space
taken up by the old answer has to be recovered by garbage collection 
(the ALS Prolog's garbage collector has not yet been extended by us to include
table space garbage recovery). As a result,
if arguments of tabled predicates are bound to structures, more
table space may be used up. 

\section{A Declarative Method for Dynamic Programming}
\label{sec:dynamic}

In the dynamic programming paradigm  the value of an optimal  solution
is recursively defined  in  terms  of  optimal     solutions  to
subproblems. Such dynamic programming definitions
can be   very tricky and  error-prone to specify due to the involvement 
of both optimization and recursion. In this section, we presents a novel, 
elegant method based on  tabled Prolog programming 
that simplifies the specification  of such dynamic programming solutions. 
With the mode-declaration scheme, there  is  no need  to  define the 
value of  an  optimal solution recursively, instead, defining  a general  
solution suffices for dynamic programming. The optimal value  as well as  
its corresponding concrete solution can be derived implicitly and 
automatically using tabled Prolog systems. 

\subsection{Dynamic Programming with TLP}
\label{sec:dp-tlp}

Dynamic   programming  algorithms  are   particularly appropriate  for
implementation    with tabled logic programming \cite{warren}. Dynamic
programming is typically used for solving optimization problems. It is
a general recursive strategy in which  optimal solution to a problem
is defined in terms of optimal  solutions to  its   subproblems.  Dynamic
programming, thus,  recursively reduces the  solution to  a problem to
repetitively  solving its  subproblems.   Therefore, for computational
efficiency it is essential that a given subproblem is solved only once
instead  of multiple times.   From   this  standpoint, tabled    logic
programming {\em dynamically} incorporates the   dynamic
programming   strategy \cite{warren}    in the    logic    programming
paradigm. TLP systems  provide implicit tabulation scheme for  dynamic
programming, ensuring that subproblems are evaluated only once.

In  spite  of the  assistance of tabled  resolution, solving practical
problems with dynamic  programming is  still not  a trivial task.  The
main step in the dynamic programming paradigm is to define the value of an
optimal  solution recursively in   terms of the  optimal solutions  to
the subproblems. This definition could  be very tricky and error-prone 
due to the interleaving of optimization and recursion. As the most widely 
used  TLP system,    XSB provides table aggregate 
predicates \cite{xsb,swift},  such  as {\tt bagMin/2}  and {\tt  bagMax/2},  
to  find the minimal   or  maximal value from tabled answers respectively.  
Those  predicates are  helpful  in finding the
optimal solutions, and  therefore in  implementing dynamic programming
algorithms.  However,  users  still have  to  define optimal solutions
explicitly, that is,  specify how the  optimal value of  a problem
is  recursively defined in terms of the optimal values of its subproblems. 
Furthermore,  the  aggregate predicates   require the  TLP   system to
collect all possible values, whether  optimal or non-optimal, into the
memo table, which could dramatically increase the amount of table space needed.

We use  the {\em  matrix-chain multiplication}   problem \cite{alg} as an
example to illustrate how tabled  logic programming can be adopted for
solving  dynamic programming problems.  A product   of  matrices is {\em  fully
parenthesized} if it is either  a single matrix  or the product of two
fully parenthesized matrix  products, surrounded by parentheses. Thus,
the matrix-chain multiplication problem    can be stated   as  follows
(detailed  description  of  this problem  can  be found  in  any major
algorithm textbook covering dynamic programming):

\begin{Problem}
\label{ex:matrix}
Given a chain  $\langle A_1, A_2, ...,  A_n  \rangle$ of $n$  matrices,
where   for $i=1,2,...,n$, matrix   $A_i$ has dimension $p_{i-1}\times
p_i$, fully  parenthesize  the product $A_1A_2...A_n$  in  a  way that
minimizes the number of scalar multiplications.
\end{Problem}

To  solve this problem via dynamic  programming, we  need to define the
cost  of an   optimal solution  recursively  in  terms of the  optimal
solutions to  subproblems.   Let $m[i,j]$ be   the  minimum number  of
scalar multiplications needed to compute  the matrix $A_{i..j}$, which
denotes a sub-chain of matrices  $A_iA_{i+1}...A_j$ for $1 \leq i \leq
j  \leq  n$. Thus, our  recursive  definition for the minimum  cost of
parenthesizing the product $A_{i..j}$ becomes

\begin{equation*}
m[i,j] = \left\{ \begin{array}{ll}
            0 & \mbox{\ \ if $i=j$,} \\
            \raisebox{-1.6ex}{$\stackrel{\textstyle \mathbf{min}}{\mathit{\scriptstyle i\leq k < j}}$}\ 
            \{m[i,k]+ m[k+1, j] + p_{i-1}p_{k}p_{j}\} & \mbox{\ \ if $i<j$.}
            \end{array}
         \right.
\end{equation*}

As shown in Example~\ref{pgm:scalar}, a tabled  Prolog coding is  given to solve
the  matrix-chain    multiplication problem.    The predicate  {\tt
scalar\_cost(PL, V, P0, Pn)} is tabled, where  {\tt
PL}, {\tt P0} and  {\tt Pn} are  given  by the  user to represent  the
dimension  sequence $[p_0, p_1,  ..., p_n]$, the first dimension $p_0$
and the last dimension $p_n$, respectively, and {\tt V} is the minimum
cost  of scalar multiplications  to  multiply $A_{1..n}$; the built-in
predicate  {\tt findall(X,G,L)} is used to   find all the instances of
{\tt X} as a list {\tt L}  such that each  instance satisfies the goal
{\tt G}; the predicate {\tt break(PL, PL1, PL2, Pk)}  is used to split
the dimension sequence  at  the point of   {\tt Pk} into two  parts to
simulate the parenthesization; and the predicate {\tt list\_min(L, V)}
finds a minimum number {\tt V} from a given list {\tt L}.

\begin{Ex}
A tabled logic program for matrix-chain multiplication problems:
\label{pgm:scalar}
\end{Ex}
\begin{quote}
{\tt 
:- table scalar\_cost/4.\\
scalar\_cost([P1, P2], 0, P1, P2).\\
scalar\_cost([P1, P2, P3 | Pr], V, P1, Pn) :-\\
\verb+    +findall(V, ( break([P1, P2, P3 | Pr], PL1, PL2, Pk),\\
\verb+                 +scalar\_cost(PL1, V1, P1, Pk),\\
\verb+                 +scalar\_cost(PL2, V2, Pk, Pn),\\
\verb+                 +V is V1 + V2 + P1 * Pk * Pn ), VL),\\
\verb+    +list\_min(VL, V). \\
break([P1, P2, P3],[P1, P2], [P2, P3], P2).\\
break([P1, P2, P3, P4|Pr],[P1, P2],[P2, P3, P4|Pr],P2).\\
break([P1, P2, P3, P4|Pr], [P1|L1], L2, Pk) :-\\
\verb+    +break([P2, P3, P4|Pr], L1, L2, Pk).
}
\end{quote}

Consider the problem  for a chain   $\langle A_1, A_2,  A_3 \rangle$ of
three matrices.  Suppose that the dimensions of the matrices are $10
\times 100$, $100  \times 5$, and $5  \times 50$, respectively. We can
give a query  {\tt :- scalar\_cost([10, 100,   5, 50], V, 10,  50)} to
find the minimum  value of its  scalar multiplications.  As  a result,
{\tt  V} is  instantiated to  $7500$,  corresponding to the   optimal
parenthesization $((A_{1}A_2)A_3)$.

Program~\ref{pgm:scalar}  shows  that the programmer   has to find the
optimal   value   by  comparing  all  possible   multiplication  costs
explicitly.   In  fact, for   a   general optimization   problem,  the
definition of  an optimal solution could  be  quite complicated due to
heterogeneous  solution  construction.  Then,  comparing  all possible
solutions explicitly to find the optimal  one can be complicated. 

\subsection{A Declarative Method based on Modes}

We present a declarative method based on the mode declaration scheme 
to separate the task of finding the optimal  solution from the  
task of specifying the general  dynamic  programming  formulation.  
Using  our   method,  the programmer is  only required to 
define {\em  what}  a general solution is,  while searching  
for the  optimal  solution is  left  to the  TLP system.
 
The mode declaration can be used  to make control of execution implicit
during  dynamic  programming,   making  the specification of
dynamic  programming problems  more declarative and elegant.  For  the
matrix-chain multiplication, instead   of   defining the cost  of   an
optimal solution, we only need to specify what the  cost for a general
solution is.   Let $m[i,j]$  be  the number of  scalar multiplications
needed to compute the matrix $A_{i..j}$ for $1 \leq  i \leq j \leq n$,
where $n$ is the total  number of matrices.  The recursive  definition
for the cost of parenthesizing $A_{i..j}$ becomes
\begin{equation*}
m[i,j] = \left\{ \begin{array}{ll}
            0 & \mbox{\ \ \ \ \ if $i=j$,} \\
            m[i,k]+ m[k+1, j] + p_{i-1}p_{k}p_{j} &  \mbox{\ \ \ \ \ if $i<j$,}
            \end{array}
         \right.
\end{equation*}
where any $k \in [i,j)$. Thus, we have the following program shown in
Example~\ref{pgm:scalar_mode}.

\begin{Ex}
A tabled logic program with  optimum mode declaration for matrix-chain
multiplication problems:
\label{pgm:scalar_mode}
\end{Ex}
\begin{quote}
{\tt 
:- table scalar\_cost(+, min, -, -).\\
scalar\_cost([P1, P2], 0, P1, P2).\\
scalar\_cost([P1, P2, P3 | Pr], V, P1, Pn) :-\\
\verb+    +break([P1, P2, P3 | Pr], PL1, PL2, Pk),\\
\verb+    +scalar\_cost(PL1, V1, P1, Pk),\\
\verb+    +scalar\_cost(PL2, V2, Pk, Pn),\\
\verb+    +V is V1 + V2 + P1 * Pk * Pn.
}
\end{quote}

\begin{figure}[htb]
\centerline{\fbox{\psfig{figure=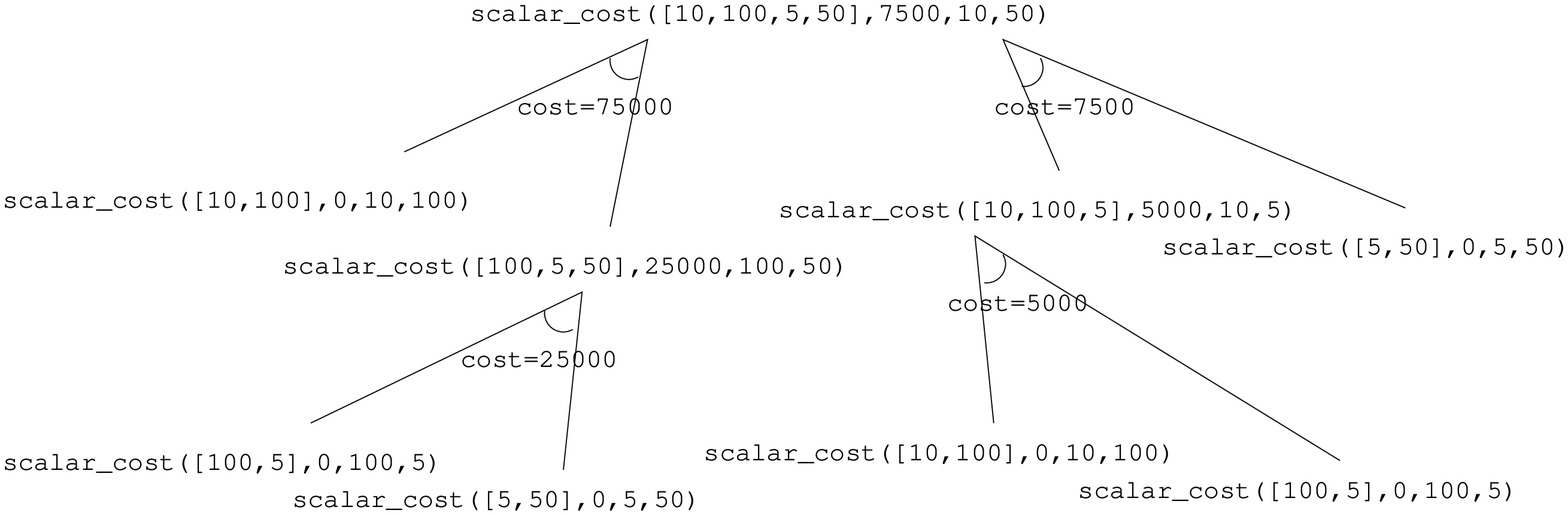,height=4.2cm,width=10.8cm}}}
\caption{Recursion tree for computing {\tt scalar\_cost([10,100,5,50],V,10,50)}}
\label{fig:matrix-mode}
\end{figure}

The mode declaration {\tt scalar\_cost(+,min,-,-)}  means that only  the
first argument (the  list of matrix dimensions)   is used for  variant
checking when an answer is generated,  and a minimum value is expected
from the  second   argument  (the  cost  of   scalar  multiplication).
Figure~\ref{fig:matrix-mode} shows a skeleton of the recursion  tree produced by the
query\\  
\centerline{\tt     :- scalar\_cost([10,100,5,50],V,10,50).}\\  
Its first tabled answer has {\tt V=75000}.  However, when the second answer {\tt
V=7500} is computed, it will automatically replace the previous answer
following  the  declared  optimum  mode. Therefore, there   is at  most one
answer for the tabled call {\tt scalar\_cost([10,100,5,50],V,10,50)} that exists in
the table at any point  in time, and it  represents the optimal  value
computed up to that point.  

To make the matrix-chain  multiplication problem complete, we  need to
construct an  optimal parenthesization  solution corresponding  to the
minimal cost   of scalar   multiplication. This construction    can  be
achieved with the  strategy  of introducing an extra non-indexed 
argument whose instantiation becomes the solution. 
The complete tabled logic program is shown below:

\begin{Ex}
A tabled logic program for the complete matrix-chain multiplication problem:
\label{pgm:scalar_mode_evid}
\end{Ex}
{\tt 
:- table scalar\_cost\_evid/5.\\
:- table\_mode scalar\_cost\_evid(+, min, -, -, -).\\
scalar\_cost\_evid([P1, P2], 0, P1, P2, (P1,P2)).\\
scalar\_cost\_evid([P1, P2, P3 | Pr], V, P1, Pn, (E1*E2)) :-\\
\verb+    +break([P1, P2, P3 | Pr], PL1, PL2, Pk),\\
\verb+    +scalar\_cost\_evid(PL1, V1, P1, Pk, E1),\\
\verb+    +scalar\_cost\_evid(PL2, V2, Pk, Pn, E2),\\
\verb+    +V is V1 + V2 + P1 * Pk * Pn.
}

\section{Experimental Results}
\label{sec:experiment}

Our experimental benchmarks  include five typical dynamic  programming
examples.  {\tt  matrix} is the  matrix-chain multiplication  problem;
{\tt lcs} is longest common subsequence problem; {\tt obst} finds
an optimal  binary search tree;  {\tt apsp}  finds the  shortest
paths for   all pairs  of nodes;    and {\tt  knap} is   the knapsack
problem.  All tests  were performed in TALS system on an Intel   Pentium 4 CPU
2.0GHz machine with 512M RAM running RedHat Linux 9.0.

\begin{table}[htb]
\centering
\begin{tabular}{c}
\begin{tabular}{|c|c|c|}
\hline\hline
{\bf Benchmarks} &  \emph{without modes} & \emph{with modes}\\
\hline\hline
{\bf matrix} & 1.99 (1.0) & 1.01 (0.51) \\
\hline
{\bf lcs} &  0.80 (1.0) &  0.37 (0.46) \\
\hline
{\bf obst}	&  0.81 (1.0) &  0.29 (0.36) \\
\hline
{\bf apsp} & 3.73 (1.0) &  2.71 (0.73) \\
\hline
{\bf knap} & 51.79 (1.0) &  35.04 (0.68) \\
\hline\hline
\end{tabular}
\smallskip
\\
(a) Without evidence construction
\\
\\
\begin{tabular}{|c|c|c|}
\hline\hline
{\bf Benchmarks} &  \emph{without modes} & \emph{with modes}\\
\hline\hline
{\bf matrix} & 2.74 (1.0) & 1.97 (0.72) \\
\hline
{\bf lcs} &  0.86 (1.0) &  0.55 (0.63) \\
\hline
{\bf obst}	&  10.58 (1.0) &  0.63 (0.06) \\
\hline
{\bf apsp} & 6.05 (1.0) &  2.85 (0.47) \\
\hline
{\bf knap} & 126.25 (1.0) &  38.56 (0.31) \\
\hline\hline
\end{tabular}
\smallskip
\\
(b) With evidence construction
\end{tabular}
\caption{Running time performance comparison in Seconds (Ratio)}
\label{tbl:time}
\end{table}

Table~\ref{tbl:time} compares the running time performance between the programs with 
and without mode declaration. The first group of benchmarks consists of programs that only 
seek the optimal value without evidence construction, while the second group
consists of programs for the same dynamic programming problems with evidence construction. 
The experimental data indicates, based on the ratios in Table~\ref{tbl:time}, 
that the programs with mode declaration consume only $6\%$ to $73\%$ of the running
time compared to corresponding programs without mode declaration.

\begin{figure}[htb]
\begin{tabular}{cc}
\includegraphics[width=2.2in,height=2.45in]{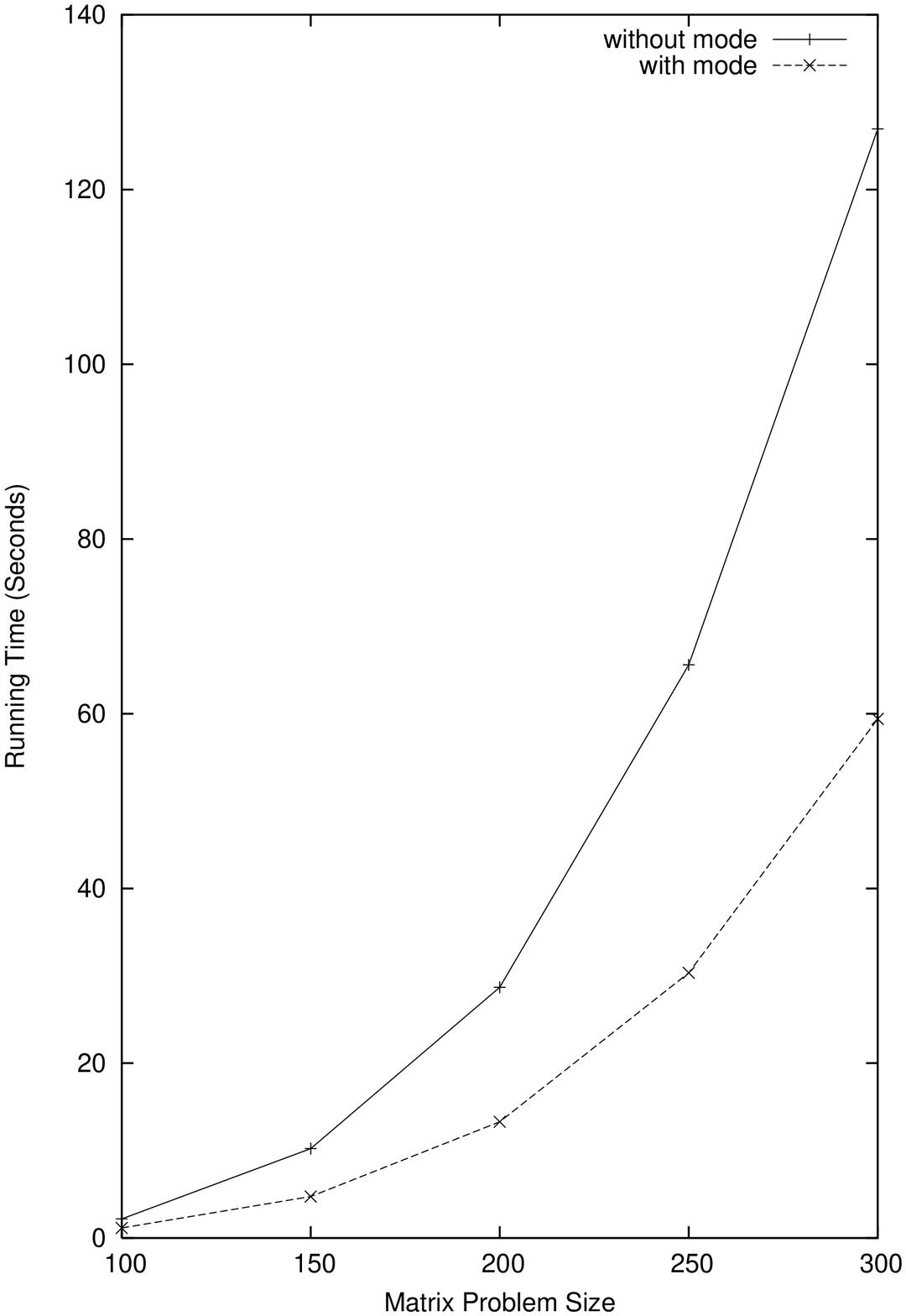}
&
\includegraphics[width=2.2in,height=2.45in]{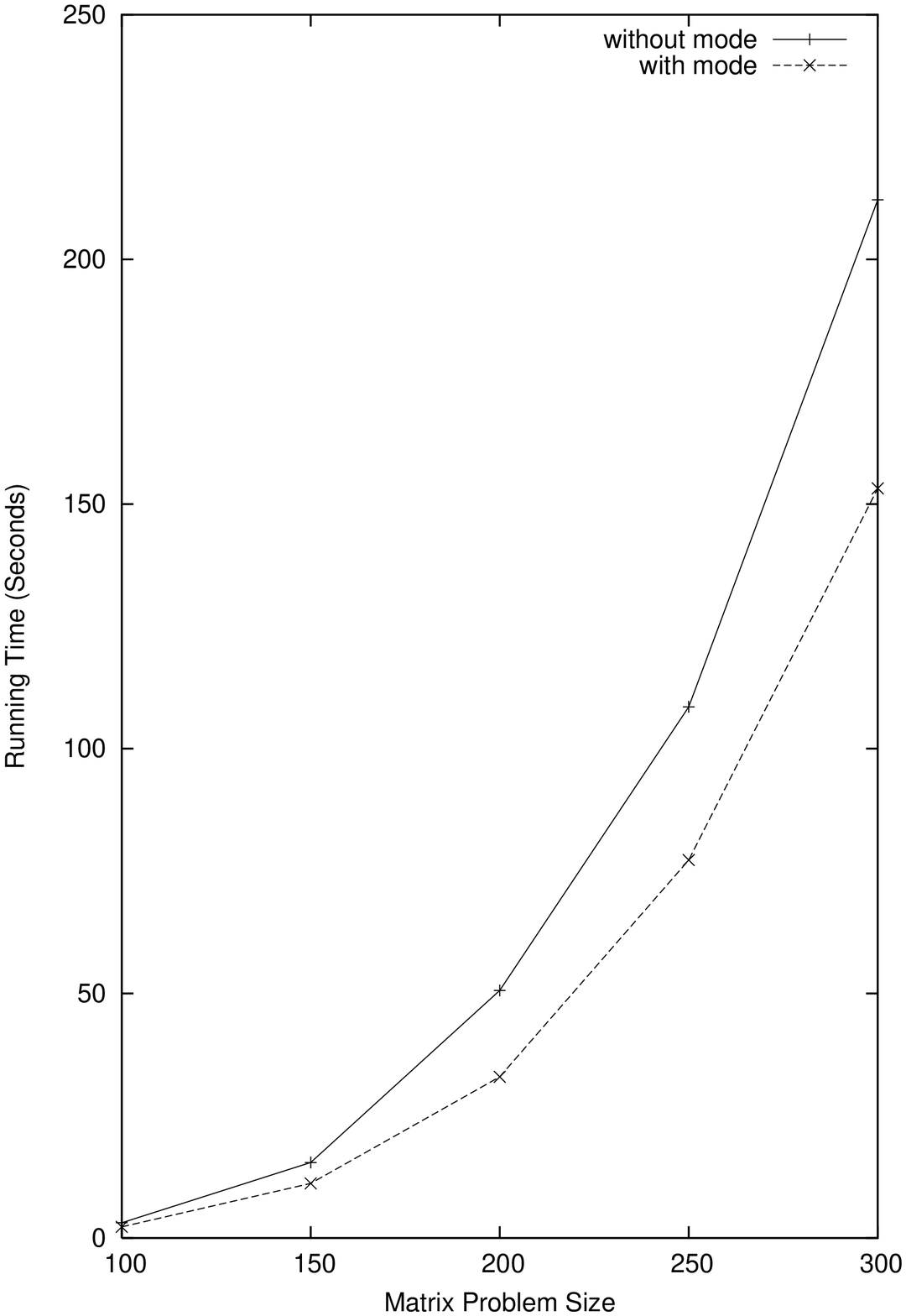}
\\
(a) Without evidence
& 
(b) With evidence
\\
\end{tabular}
\caption{Time Performance of Matrix-chain Multiplication}
\label{fig:perm}
\end{figure}

Figure~\ref{fig:perm} shows the timing information against different 
input sizes for matrix-chain multiplication problems. Notice that the numbers on X-axis
represent the total number of matrices to be multiplied, and the numbers
on Y-axis represent the running time with seconds. Whether without evidence
construction (Figure~\ref{fig:perm}(a)) or with evidence construction (Figure~\ref{fig:perm}(b)),
the graphs indicate that the timings of the programs with mode
are consistently better than those without mode declaration. 

The efficiency of these programs is mainly credited to two factors.
First, tabled Prolog systems with mode declaration provides a concise but easy-to-use
interface for dynamic programming, and it does not introduce any major 
overhead; mode declarations are flexible and powerful means of supporting meta-level 
manipulation of fixed-points; and, mode functionality is implemented at the system level
rather than at the Prolog programming level. 
Second, tabled answers can be more efficiently organized due to the mode declaration.
Indeed, if an indexed argument is instantiated in advance 
before a tabled goal is called, variant checking on
this indexed argument can be avoided since its value is the same for all the answers;
furthermore, it is not necessary to record the pre-instantiated value 
with each tabled answer because the same value has already been stored 
in the tabled call entry. These optimizations lead to considerable 
improvement in run-time performance .

The main disadvantage of our scheme w.r.t. efficiency is the need for 
frequent retrieval or replacement of tabled answers. This is 
because the optimized answer is dynamically selected 
by comparing it with old tabled answers according to the modes. 
The retrieval of a tabled answer for comparison incurs time overhead
since information about each argument of the answer needs to be retrieved from the table.
For replacing a tabled answer in the current TALS system, if a tabled subgoal only
involves numerals as arguments, then the tabled answer will be completely
replaced if necessary. If the arguments involve structures, however,
then the answer will be updated by a link to the new answer. Space
taken up by the old answer has to be recovered by garbage collection later. 
As a result, the retrieval and replacement of tabled answers
incurs performance overhead, especially space overhead as explained further below. 
The overhead will be minimal if the first
tabled answer for each tabled call is optimal.

We compare the running space performance 
between the programs with and without mode declaration in Table~\ref{tbl:space}. 
For benchmarks without evidence construction, 
our experiments indicate that with mode declaration, programs consumes only
$7\%$ to $73\%$ of the space compared to those without mode declaration. 
With evidence construction included,
space performance can be better or worse depending on the problem. 
For the {\tt matrix} and {\tt obst} problems that try to find the optimal 
binary tree structure, the programs without mode declaration
generate all possible answers and then only table the optimal one, 
while the programs with mode declaration and implicit aggregation
generate all possible 
answers and selectively table the better answers until the optimal one is
found. In the latter case, some non-optimal answers may be replaced
in the table during the computation, however, the space taken by those
old answers, including tree structures, cannot be recovered immediately. 
If the optimal answer happens to be
the first tabled answer,
then no other un-optimal answers will be recorded.
This is the reason why the benchmarks {\tt matrix} and {\tt obst}
(with evidence construction) with mode declaration 
take more space than those without mode, as shown in Table~\ref{tbl:space}.

\begin{table}[htb]
\centering
\begin{tabular}{c}
\begin{tabular}{|c|c|c|}
\hline\hline
{\bf Benchmarks} &  \emph{without modes} & \emph{with modes}\\
\hline\hline
{\bf matrix} & 4.98 (1.0) & 0.93 (0.19) \\
\hline
{\bf lcs} &  78.75 (1.0) &  23.57 (0.30) \\
\hline
{\bf obst}	&  2.65 (1.0) &  0.63 (0.24) \\
\hline
{\bf apsp} & 20.17 (1.0) &  14.68 (0.73) \\
\hline
{\bf knap} & 222.65 (1.0) &  16.29 (0.07) \\
\hline\hline
\end{tabular}
\smallskip
\\
(a) Without evidence construction
\\
\\
\begin{tabular}{|c|c|c|}
\hline\hline
{\bf Benchmarks} &  \emph{without modes} & \emph{with modes}\\
\hline\hline
{\bf matrix} & 9.22 (1.0) & 12.01 (1.30) \\
\hline
{\bf lcs} &  92.99 (1.0) &  44.55 (0.48) \\
\hline
{\bf obst}	&  4.44 (1.0) &  19.37 (4.36) \\
\hline
{\bf apsp} & 29.90 (1.0) &  21.60 (0.72) \\
\hline
{\bf knap} & 399.95 (1.0) &  306.56 (0.77) \\
\hline\hline
\end{tabular}
\smallskip
\\
(b) With evidence construction
\end{tabular}
\caption{Running space comparison: Megabytes (Ratio)}
\label{tbl:space}
\end{table}

\section{Conclusions}
\label{sec:conclusion}

We presented a new tabled resolution scheme based on
{\em dynamic reordering of alternatives (DRA)} 
for efficient fixed-point computation. 
It works with a single SLD-similar tree without suspension of any goals,
and its operational semantics mimics the semi-naive bottom-up computation.
Our scheme can be easily implemented on top of an existing
Prolog system without modifying the kernel of the WAM engine in any major way.
We were able to implement it on top of an existing Prolog engine (ALS Prolog) 
in a few man-months of work. Performance evaluation of our implementation shows that
it is comparable in performance to well-engineered tabled systems
such as XSB, yet it is considerably easier to implement. 

A  mode  declaration scheme for  tabled predicates  was also introduced 
in  TLP systems to aggregate information dynamically  into the  table.
These modes provide a declarative method to reduce a fixed-point from a large
 solution set to a preferred small one, or from an infeasible infinite
 solution set to a finite one. The mode declaration classifies arguments of   tabled predicates as
either indexed or non-indexed. As a result, (i) a tabled predicate can
be regarded as a function in which non-indexed arguments (outputs) are
uniquely  defined by the    indexed arguments (inputs);  (ii)  concise
explanation    for tabled   answers   can be    easily  constructed in
non-indexed (output)    arguments;  (iii) the    efficiency  of tabled
resolution can  be improved since  only indexed arguments are involved
in variant checking; and (iv) the  non-indexed arguments of a tabled
predicate can be further qualified with an aggregate mode such that an
optimal value can be sought without explicit coding of the comparison.

This new mode declaration  scheme, coupled with recursion, provides an
elegant method for specifying dynamic programming problems: there is
no need to  define the value  of an optimal solution recursively,
instead,  defining  the value of a  general  solution is  enough.  The
optimal  value,  as well as  its   associated  solution, is  obtained
automatically by the TLP systems. This new scheme has been implemented in the authors' 
TALS system with very encouraging results. 

\section*{Acknowledgments}

We are grateful to David S. Warren, C. R. Ramakrishnan, Bart Demoen,
Kostis Sagonas and Neng-Fa Zhou for general discussions about tabled  logic
programming and to Peter Stuckey for discussion on work presented in
this paper.

\bibliographystyle{acmtrans}

\end{document}